# Study of electrical conductivity of the coatings of bimetallic Au-Ag nanoparticles

A.G. Sivakov[1], R.P. Yavetskiy[2], N.A. Matveevskaya[2], T.G. Beynik[2], A.V. Tolmachev[2], S.I. Bondarenko[1], A.S. Pokhila[1], A.V. Krevsun[1], V.P. Koverya[1], A.S. Garbuz[1]

[1]*B. Verkin Institute for Low Temperature Physics and Engineering, National Academy of Sciences of Ukraine, Nauky Ave., 47, 61103 Kharkiv, Ukraine*
[2]*Institute for Single Crystals, National Academy of Sciences of Ukraine Nauky Ave., 60, Kharkiv, 61178, Ukraine*

e-mail: bondarenko@ilt.kharkov.ua

The "metallic" temperature variation of the electrical resistivity of chemical nano gold-silver coatings with a silver content of 6.6% at. up to 13.1% at. determined in the temperature range 4.2–300 K. Features of low-temperature measurements of the resistance of nano-coatings were found and explained. The possibility of the appearance of superconductivity in them is discussed.

## 1. Introduction

In July 2018, a message appeared from Indian scientists about the development and properties of a new superconducting compound in the form of a chemical coating of nanoparticles (NPs) of gold with silver [1]. The temperature dependences of the resistance of manufactured coatings, on which the linear change of resistance typical for metals is replaced by a sharp drop to zero in the temperature range of 236 K (–37 °C) characteristic of the superconducting transition, are presented in this message. Measurement of the magnetic permeability of the coating samples showed the presence of a diamagnetic response below the transition temperature, which indicates the presence of the Meissner effect. In addition, an external magnetic field of 0.1 to 3 T biased the beginning of the transition to lower temperatures.

The combination of these experimental results is usually sufficient evidence of the existence of superconductivity. It was also reported that one of the samples had a transition temperature of 320 K (+47 °C), which meant that room temperature superconductivity (PCB) was achieved for the first time in the history of superconductivity with normal external atmospheric pressure. Attempts to test the possibility of the existence of such a compound [2,3] after the specified message were made. In [2], nanoscale films consisting of alternating layers of gold and silver were made and studied. It should be noted that the vacuum thermal method of sputtering gold and silver, applied in [2], is very far from the chemical synthesis of these components in [1]. As a result, superconductivity has not been found out. The technique for measuring the temperature dependences of coatings of NPs using the four-probe method was theoretically analyzed in [3]. It was shown that with certain properties of the coating and the location of the potential and current leads from the sample, it is possible to obtain a zero or close to zero value of the potential difference on the sample due to the non-uniform distribution of the measuring current between the NP, even in the absence of NP superconductivity. Zero-resistance of a sample can be mistaken for the appearance of its superconductivity. At the same time, the author [3] admits that the presence of the Meissner effect in the samples of [1] may be evidence of the existence of superconductivity in them (even if the result of transport measurements is erroneously explained).

The aim of our work was to experimentally test the transport and some magnetic properties of gold–silver (Au–Ag) nanocoatings prepared by the chemical method.

**2. Samples preparation method**

To obtain the coatings based on "star"-shape bimetallic nanoparticles (NP) Au-Ag we used chloroauric acid $HAuCl_4·3H_2O$ (≥99.9%); trisodium citrate $Na_3C_6H_5O_7$ (≥98.0%); hydrochloric acid HCl (~38.0%); silver nitrate $AgNO_3$ (≥99.8%); ascorbic acid $C_6H_8O_6$ (≥99.0%); (3-aminopropyl)triethoxysilane (APTES) $C_9H_{23}NO_3Si$ (≥99.0%); sulfuric acid $H_2SO_4$ (~98.0%); hydrogen peroxide $H_2O_2$ (~38.0%); ethyl hydroxide $C_2H_6O$ ~ 96 % were purchased from Sigma-Aldrich, Germany. All chemical reagents were used as received without further purification.

Aqueous NP colloidal solutions were obtained by a two-step method on Au nucleating seeds by reduction of chloroauric acid silver nitrate by ascorbic acid. In [4] we describe the NP Au-Ag synthesis in detail. The Au-Ag NP films were obtained by self-assembly method [5, 6], based on multistage NP absorption from the colloidal solution onto the substrate. In order to improve the first layer NP absorption, the substrate surface was chemically modified by APTES molecules [6]. After that substrates were immersed into a fresh prepared colloidal solution of NP and soaked for 24 hours in the temperature of 20 ± 2 °C to form Au-Ag NP coating, then repeatedly flushed out and air-cured. To increase the surface thickness the second and third cycles of NP absorption onto the substrate were processed analogously (the thickness of the NP layer after the third cycle is 800 nm).

Morphology and structure of Au-Ag NP were studied by transmission electron microscopy (TEM) with the use of electron microscope TEM-125 (Selmi, Ukraine) with accelerating voltage of 100 kV. The statistical analysis of NP sizes obtained by the electron microscopy was processed, and the NP size distribution histograms were constructed. For every NP size obtained in the process the diameter of at least 300 particles was measured. Morphology and surface profile of NP coatings were studied by the high-resolution scanning electron microscope (HR-SEM) (Hitachi S-5500, Hitachi High-Technologies Corporation, Japan) and the atomic force microscope (AFM) (SPM-9600, Shimadzu Corp., Japan) in non-contact mode with frequency of 0.2 Hz using PPP-NCHR cantilever. Sample composition was analyzed by the scanning electron microscope JSM-6390LV with the X-ray chemical analysis system INCA Energy 350. Localness of energy dispersive X-ray analysis was ~1μm, concentration accuracy was 1-5 rel. %. Samples homogeneity was checked applying the assaying method point-to-point.

Stable colloidal solutions of branched Au-Ag NP with the average particle size of 55-73 nm were obtained. Silver ions presence is an important condition for forming and growth of sharp branches of "star"-shaped NP along certain crystallographic faces of Au nucleating seeds. We suppose that silver absorbs onto the surfaces of seeds faces with the largest surface energy, forms monolayers [7, 8], selectively stabilizes faces {110}, {310}, {720} and prevents further growth of Au on the surface of these faces. Anisotropic growth of NP leads to the forming of side branches on faces with lower surface energy. After the stabilization of faces {110} Au absorbs on faces {111}, that leads to the forming of gold branches. Increase of the silver ions concentration in the growth solution leads to an increase of the average NP size. The concentration of silver in Au-Ag NP of different size is represented in Table 1.

Table 1. Concentration of silver in Au-Ag NP.

| Sample number | Average NP size, nm | Ag concentration, at.% |
|---|---|---|
| 1. | 55 | 6.6 ± 0,3 |
| 2. | 68 | 10.8 ± 0,3 |
| 3. | 73 | 13.1 ± 0,3 |

In Fig. 1 electron microscope image and size distribution histograms of Au-Ag NP with branched morphology are represented. NP with average size of 55 and 68 nm have ~ 20% size dispersion (fig. 1 a, b), and NP with average particle size of 73 nm have ~30% size dispersion (fig. 1 c).

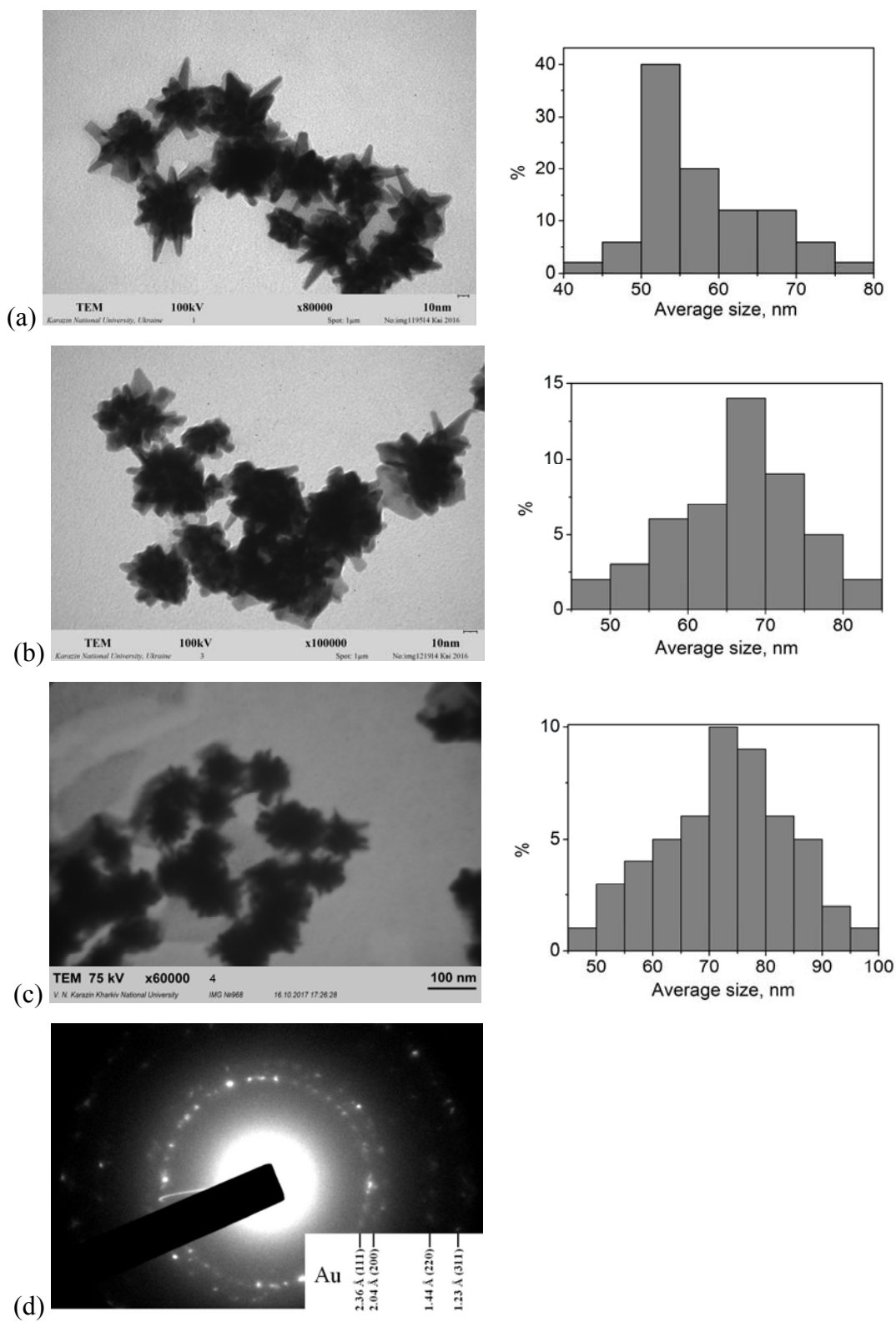

Fig. 1. TEM of branched Au-Ag NP with average size: (a) 55 nm, (b) 68 nm, (b) 73 nm and the corresponding NP size distribution histograms; (d) electric micro diffraction of NP.

Obtained branched Au-Ag NP are nanocrystals, and the existence of distinctive reflexes, corresponding to the faces {111}, {200}, {220}, {311}, indicate a face-centered cubic lattice of Au-Ag NP (fig.1 d).

Using self-assembly method, stable films with different Ag concentration were obtained (fig. 2). Produced films are characterized by a framework (arc) structure defined by a branched shape of Au-Ag NP. Analysis of films surfaces by the AFM method showed that the average thickness of NP layer is 800 nm (fig. 3).

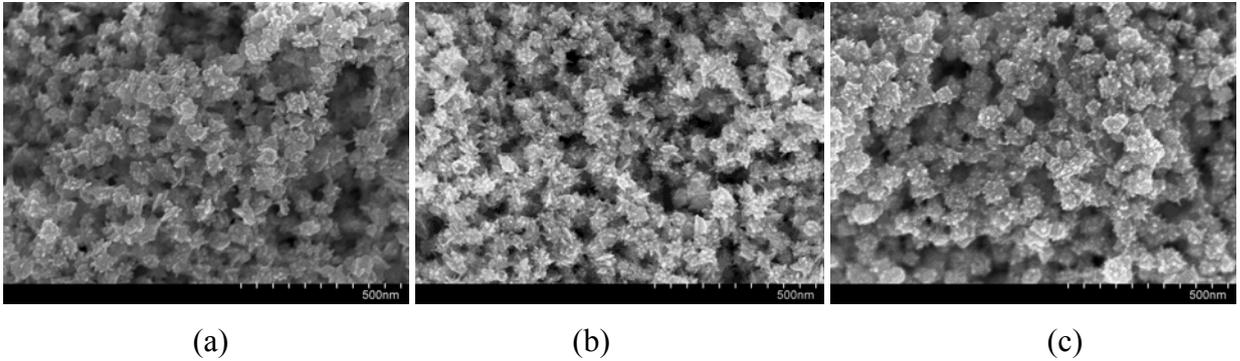

(a)          (b)          (c)

Fig. 2. TEM of films surfaces of NP with different Ag concentration: (a) 6.6 at. %, (b) 10.8 at. %, (c) 13.1 at. %.

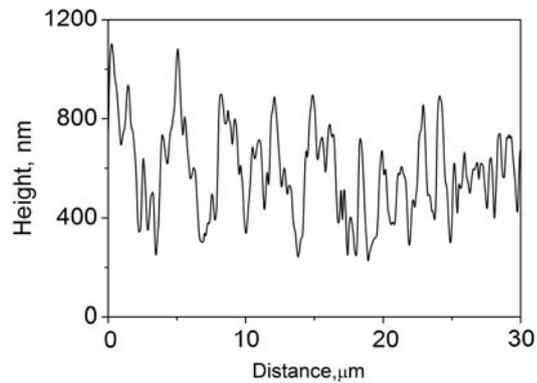

Fig. 3. Surface profile of NP coating with Ag concentration of 6.6 at. %, layer thickness is 800 nm.

**3. Method of electrical conductivity investigation**

The study of the electrical conductivity of samples with a coating of NP Au–Ag on glass consisted of measuring their temperature dependences of resistance using four probe methods in the temperature range 300–4.2 K. The connection of current and potential wires with the sample was carried out in one of two ways. In the first method, strips of indium-tin alloy or strips of indium melted in air were deposited in vacuum across the sample on the coating. Gold wires with a diameter of 0.05 mm were glued to the contact areas of the strips with silver paste. In the second method, the current leads were glued with silver paste directly on the edge of the sample. The typical length of the samples was 5 mm with a width of 1–2 mm. The distance between potential contacts ranged from 0.1 to 1 mm. Measurement of the resistance ($R$) of the sample as a function of temperature ($T$) was carried out automatically by recording the volt-ampere characteristics ($I$–$V$) of the sample at different temperatures and converting them depending on $R(T)$. The maximum constant measuring current in various experiments was from 50 to 800 µA. $I$–$V$ and and $R(T)$ dependences were linear when using the first method of formation of current and potential leads in the whole range of measuring currents (in detail look more low). The magnetic properties of the samples were controlled in two ways. In the first of them, a permanent magnet was located above the surface of the sample, creating in the coating area a magnetic field of 0.05 T. The temperature dependence of the resistance of the sample in the temperature range of 4.2–300K was then recorded. Then the dependencies obtained with and without a magnet were compared. In the second method, the coated substrate could be placed on a flat coil with a diameter close to the size of the coating. After that, measurement and comparison of the inductance value of the coil without a coated substrate and when it was installed on the coil was made. The inductance was measured using a high-frequency digital inductance meter type E7-12 at a frequency of 1 MHz. The resolution of the meter was 100 nH.

**4. Research results and discussion**

Typical temperature dependences of the resistance $R(T)$ of samples with different silver contents in the temperature ranges from room temperature to 100 K are shown in Fig.4.

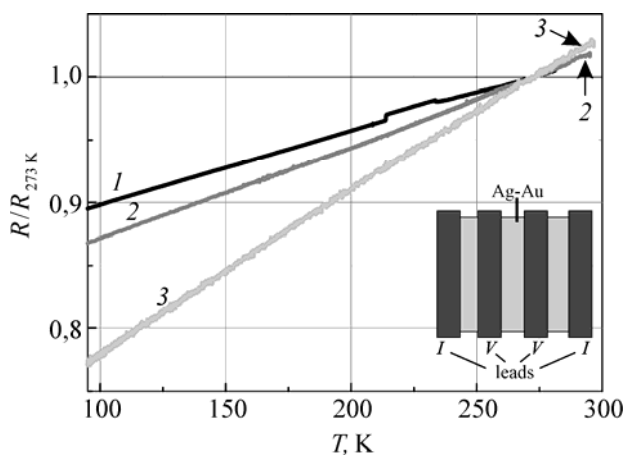

Fig. 4. Dependencies $R/R_{273\ K}(T)$ of the resistance of the samples with different silver contents, reduced to $T = 273$ K: *1* — 6.6 % Ag (black line), *2* — 10.8% Ag (dark gray line), *3* — 13.1 % Ag (light gray line). The inset shows the layout of the contacts on the samples.

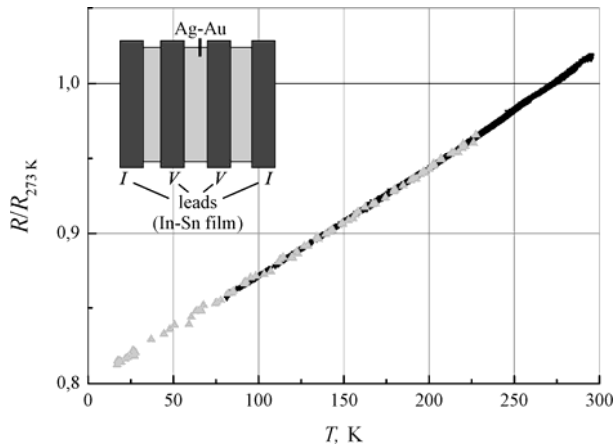

Fig. 5 Dependence of the relative resistance $R/R_{273\,K}(T)$ reduced to $T$ = 273 K of sample No. 2 (see Fig. 4), as a function of temperature, taken at intervals of 2 weeks (black line — 1st experiment, gray triangles — 2nd experiment). Contacts — In-Sn strips. The inset shows the layout of the contacts on the sample.

The common property of all curves is the almost linear ("metallic") dependence of the resistance $R(T)$ in a wide temperature range with different values of the temperature coefficient of resistance (TCR). It is seen that the TCR of the samples increases with increasing percentage of silver. It is established that TCR of coatings is 2-3 times less than TCR of pure gold and silver. All these dependences are obtained using the first method of forming current and potential contacts. InSn melt strips applied to coatings across the direction of the transport current were used as potential and current contacts. Reproducible results with multiple repetitions of the cooling-heating could be obtained only with this method of forming contacts. The dependences $R(T)$ for sample No. 2, taken at intervals of two weeks, are presented in Fig. 5. Reproducible results with multiple repetitions of the cooling-heating could be obtained only with this method of forming contacts.

The resistance of the samples may increase several times after they are stored in air. The investigated coatings were not protected from the moisture of the surrounding air. Optical studies of samples showed that they are very heterogeneous in area and have poor adhesion to glass. NP of the coatings have poor contact not only with the substrate, but also with each other. This means that the coating is a heterogeneous in conductivity and loose in the volume of the grid of contacting with each other NR. Photomicrographs in Fig. 2, 3 confirm this idea of the electrical properties of coatings. Paths of current flow in such coatings are percaler. Contacts between NPs can change and even break, changing the configuration and number of parallel current paths with decreasing temperature due to different thermal expansion coefficient of glass and coating. As a rule, these features are more often observed with the second method of connecting current leads.

The dependence, similar to a sharply non-linear temperature dependence of resistance with a superconducting transition at a temperature of 230 K, is shown in Fig. 6.

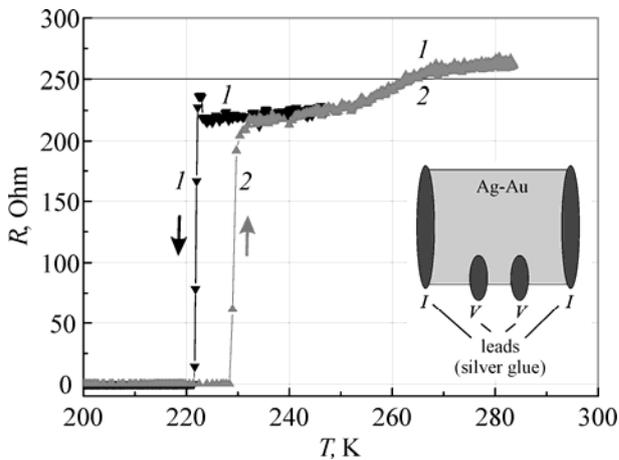

Fig. 6. Dependences *R*(*T*) of sample resistance with Ag concentration 6.6% during cooling (black triangles) and heating (gray triangles) obtained using the second method of forming contacts with a sample similar to that used by the authors of [1]. Contacts — silver paste on the edge of the sample. The inset shows the layout of contacts.

For this sample, potential leads were located on the edge of the sample. A drop in resistance to zero does not occur and non-linearities are preserved during thermal cycling, if potential contacts are applied to the same sample that intersect the entire width of the sample (inset in Fig. 7). Fig. 7 shows the dependences *R*(*T*) with three cycles of cooling and heating.

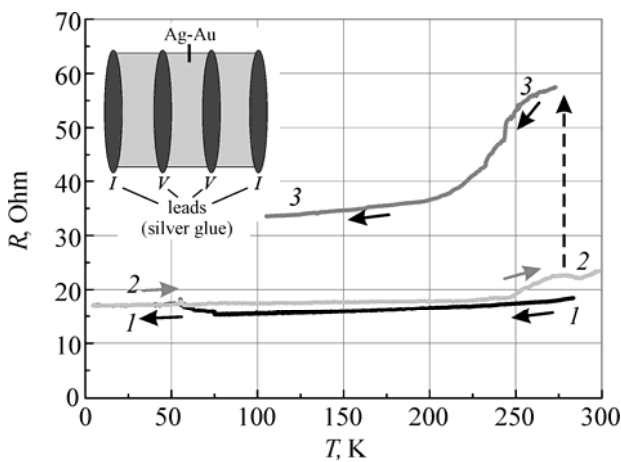

Fig. 7. Dependences *R*(*T*) of sample resistance with Ag concentration of 10.86 % during cooling (black line), heating (light gray line) and re-cooling after 1.5 hours (dark gray line). The layout of the contacts on the sample is shown in the inset, contacts of silver paste.

It can be seen that with an increase in the number of records $R(T)$, the absolute value of the resistance increases, and the nonlinearity increases. However, the independence of all curves from the magnetic field and the absence of a decrease in the inductance of the measuring coil due to the expected diamagnetic effect on the inductance of the superconductivity of the coating do not confirm its presence. Thus, a nonlinear temperature decrease in resistance, in particular, to zero on one of the Au-Ag coating samples, may not be evidence of high-temperature superconductivity in it. At the same time, the results we obtained (Fig. 6) is an experimental proof of the assumption made in [3] that the shape of the experimental $R(T)$ curves may not always be a criterion for the appearance of a superconducting transition. In the case of Fig. 6, there is a flow of the measuring current through the areas of the coating with a significantly lower local resistance than in the areas adjacent to the location of the potential contacts. As a result, these pins are actually isolated from the measuring current and there is no voltage on them. This can be perceived as the resistance of the sample to zero and the emergence of superconductivity. In fact, the non-linear portions of the curves in Fig. 6,7 can be explained by the non-linear dependence of the degree of isolation of potential contacts on the measuring current with a change in temperature. The nonlinear dependences $R(T)$ can be explained by the change in the mechanical stresses in the coating as the temperature changes, which causes a change in the total electrical conductivity of the contacts between the NP of the coating. The potential leads completely overlap the measuring current paths with the first method of measuring the $R(T)$ dependences, which makes it impossible to disconnect the potential leads from the transport current. As a result, the dependences $R(T)$ in most cases do not have non-linear sections. Weak non-linearities were observed in rare cases. Their appearance can be explained by the same processes of thermal compression (during cooling) and expansion (during heating) of a loose coating. This causes an irreversible decrease or increase in contact resistance between the NP of the coating located in the area between the potential contacts.

Thus, the dependences obtained with the correct arrangement of potential contacts (Fig. 4.5) in combination with the absence of the influence of the magnetic field on the resistance of the samples and the absence of the diamagnetic effect suggest that in the entire temperature range of the experiment the superconductivity in this type of NP of Au- Ag is missing.

Regarding the nature of the electrical conductivity of samples, the following can be noted. The change in the resistance $R$, linear with decreasing temperature, indicates their metallic conductivity. In this regard, it can be assumed that the total resistance ($R$) of the sample consists of the resistance ($R_1$) of the NP itself and the resistance ($R_2$) of the metal point contacts between them. The contribution to the conductivity of possible semiconductor interlayers between the granules is unlikely.

The following model technique is used to estimate the value of $R_1$. A model sample is made up of a complete set of substance granules. The mass of the model sample is equal to the total mass NP of the real sample. These NPs are laid so tightly that they are a solid coating without contact between the granules. Resistance $R_1$ can be calculated if the mass ($m$) of the sample coverage, the length ($l$) and width ($w$) of the sample, the specific gravity ($\sigma$) and the specific resistance ($\rho$) NP of the material are known. Our coating is rectangular in shape from densely packed NPs with dimensions in the plan of $l \times w$ (0.5 × 5 mm) and with thickness $t$. In our case $\sigma \approx \sigma_{Au} = 19.3$ g/cm$^3$, $\rho \approx \rho_{Au} \approx 10^{-6}$ Ohm·cm, $m \approx 0.001$ mg. The value of $R_1$ is calculated from two ratios:

$$R_1 \approx \rho[l/(wt)], \tag{1}$$

$$m \approx \sigma \times l \times w \times t. \tag{2}$$

From (1) and (2) we get:

$$R_1 = \frac{\rho \sigma l^2}{m}. \tag{3}$$

For the above numerical parameters from formula (3), we obtain $R_1 \approx 6 \times 10^{-2}$ Ohm·s. From a comparison of the resistance values (from 10 to 100 Ohm·s) experimentally measured at room temperature and the value of $R_1$, it follows that the contact resistance between NPs makes the main contribution to the resistance of this sample. In this case, the average specific resistance of the coating of nanoparticles ranges from $3 \times 10^{-5}$ to $3 \times 10^{-4}$ Ohms·cm. The thickness ($t$) of the model sample, based on the ratio of (2), is about 20 nm. If we take into account the thickness of the investigated coatings (800 nm), it can be obtained that its specific density is 40 times less than the density of its granules, i. e. half the water density. This also indicates a very large number of small-sized electrical contacts between the coating nanoparticles.

## 5. Conclusions

The electrical conductivity of nanoparticle (NP) Au-Ag coatings in the form of "stars" in the temperature range 4.2–300K was first determined by measuring the temperature dependence of their electrical resistivity.

The coating samples were obtained by chemical deposition of an Au–Ag colloid on glass substrates. The electrical resistance was determined by the four probe method. It is established that the temperature dependence of the coatings studied has a metallic character, i. e. the resistance decreases linearly with decreasing temperature up to the minimum temperature of the experiment (4.2 K).

The complexity of the cryogenic measurements of the electrical resistance of these samples is due to the weak and non-uniform adhesion of the coating to the substrate, as well as the low and non-uniform density of the coating. The most repeated values of electrical conductivity of samples can only be obtained using the current and potential contacts developed with the sample in the form of either deposited in vacuum over a coating of film strips of In-Sn alloy with a length equal to the width of the sample or similar strips of melted indium. False and most often non-reproducible temperature dependences of the resistance of samples from NP, which coincide in shape with typical resistive transitions to the superconducting state, can arise in the case of using traditional electrical contacts located along the periphery of the coating. In particular, one can observe the occurrence of a false zero resistance at a temperature close to room temperature. Occurrence of false temperature dependences and them non repetition from one temperature cycle to another it is possible to explain irreversible thermomechanical movings of the nanopatricles of the coatings on the substrate surface, causing casual changes of electric contacts between them. This is confirmed by an estimate of the contributions to the total resistance of the sample to the resistance of the "stars" substance and the contacts between them. The last one is predominant. The same feature of coating resistance explains their very low density.

The temperature dependences of the resistance of the "star" coatings of the selected composition, obtained using the optimal design of current and potential contacts with the sample, and the negative observation of the Meissner effect indicate the absence of superconductivity in the coatings studied.

Further search for superconductivity in such a compound is advisable to carry out using additional measurement techniques and with increasing silver content in the coating, since the investigated coatings had a silver content of up to 13 %, and in [1] superconductivity was obtained with a silver content of 15 to 20 %.

References
1. D. K. Thapa, A. Pandey, Evidence for Superconductivity at Ambient Temperature and Pressure in Nanostructures, arXiv:1807.08572 [cond-mat.supr-con].
2. A. Biswas, S. Parmar, A. Jana, R. J. Chaudhary, S. Ogale, Absence of superconductivity in pulsed laser deposited Au/Ag modulated nanostructured thin films, arXiv:1808.10699 [cond-mat.mtrl-sci].

3. D. Pekker, J. Levy, A comment on percolation and signatures of superconductivity in Au/Ag nanostructures, arXiv:1808.05871 [cond-mat.supr-con].

4. T.G. Beynik, N.A. Matveevskaya, M.V. Dobrotvorskaya, A.S. Garbuz, D.Yu. Kosyanov, V.I. Vovna, A.A. Vornovskikh, S.I. Bogatyrenko, Synthesis and characterization of branched gold nanoparticles, Functional Materials, 2017, V. 24, №1, P. 21-25.

5. O. Bibikova, J. Haas, A.I. López-Lorente, A. Popov, M. Kinnunen, I. Meglinski, B. Mizaikoff, Towards enhanced optical sensor performance: SEIRA and SERS with plasmonic nanostars, *Analyst*, 2017, 142(6), 951-958.

6. T.G. Beynik, N.A. Matveevskaya, D.Yu. Kosyanov, A.A. Vornovskikh, V.G. Kuryavyi, V.O. Yukhymchuk, O.M. Hreshchuk, Nanostructured films based on branched gold particles, *J. Nanophoton.*, 12(3), 036001 (2018).

7. P. Ndokoye, X. Li, Q. Zhao, T. Li, M.O. Tade, S. Liu, *J. Colloid Interface Sci.*, 462:341 (2016).

8. T. G. Beynik et al., Fabrication and properties of gold nanostars and film structures based on them, *Nanosist. Nanomater. Nanotekhnol.* 15(3), 417–429 (2017).